\documentclass[aps,
superscriptaddress]{revtex4}
\usepackage[intlimits]{amsmath}
\usepackage{amsfonts}
\usepackage{graphics}
\usepackage[usenames]{color}
\usepackage{epstopdf,epsfig}
\usepackage{subfigure}
\usepackage{indentfirst}
\usepackage{graphicx}% Include figure files
\usepackage{amsmath}
\usepackage{amsfonts}
\newtheorem{lem}{Lemma}

\newcommand{\e}{\varepsilon}
\newcommand{\bb}[1]{\mathbb{ #1 }}

\newcommand{\du}[1]{\frac{\delta #1}{\delta u}}

\newcommand{\beq}{\begin{equation}}
\newcommand{\eeq}{\end{equation}}
\newcommand{\de}{\partial}

\newtheorem{exa}{Example}

\newcommand{\sign}[1]{\mbox{ sign}{(#1)}}
\newcommand{\Hi}{{\cal{H}}}

\def\Xint#1{\mathchoice
   {\XXint\displaystyle\textstyle{#1}}%
   {\XXint\textstyle\scriptstyle{#1}}%
   {\XXint\scriptstyle\scriptscriptstyle{#1}}%
   {\XXint\scriptscriptstyle\scriptscriptstyle{#1}}%
   \!\int}
\def\XXint#1#2#3{{\setbox0=\hbox{$#1{#2#3}{\int}$}
     \vcenter{\hbox{$#2#3$}}\kern-.5\wd0}}
\def\dashint{\Xint-}

\begin{document}

\title{Critical behaviour for scalar nonlinear waves}
\author{Davide Masoero}\email{dmasoero@gmail.com}\affiliation{Grupo de F\'isica Matem\'atica da Universidade de Lisboa}
\author{Andrea Raimondo}\email{andrea.raimondo@unimib.it}\affiliation{Universit\`a degli Studi di Milano-Bicocca - Dipartimento di Matematica e Applicazioni}
\author{Pedro R. S. Antunes}\email{prsantunes@gmail.com}\affiliation{Grupo de F\'isica Matem\'atica da Universidade de Lisboa} \affiliation{Departamento de Matem\'{a}tica,
Universidade Lus\'{o}fona de Humanidades e Tecnologias}

\begin{abstract}
\noindent
Abstract: In the long-wave regime, nonlinear waves may undergo a phase transition from a smooth to a fast oscillatory behaviour. We study this phenomenon,
commonly known as dispersive shock, in the light of Dubrovin's universality conjecture \cite{du06,duel12}, and we argue that the transition
can be described by a special solution of a model universal partial differential equation. This universal solution is constructed  by means
of a string equation. We provide a classification of universality classes and the explicit description of the transition by
means of special functions, extending Dubrovin's universality conjecture to a wider class of equations. In particular, we show
that Benjamin-Ono equation belongs to a novel universality class with respect to the ones known in the literature, and we compute
its string equation exactly.
We describe our results using the language of statistical mechanics, showing that dispersive shocks share
many features of the tri-critical point in statistical systems, and building a dictionary between nonlinear waves and statistical mechanics.
\\


\end{abstract}

\maketitle

\section{Introduction} 

Dispersive shock waves in 1+1 dimensions are a class of shock waves recently observed in a variety of physical situations in
which the media are dispersive or not-strictly diffusive. Examples include plasma physics \cite{taylor70}, Bose Einstein condensates \cite{chang08},
nonlinear optics \cite{jwf07,cfprt09} and hydrodynamics  \cite{johnson97, zachary01, bettelheim05, konop13}. Through the shock, the waves experience an
abrupt phase transition from a regular to a rapidly-oscillatory behavior, and the transition has been conjectured to be universal, depending only on some
general properties of its underlying model  as a partial differential equation. The universality classes observed so far correspond to the class of
scalar dispersive waves (Korteweg-de Vries universality class, \cite{du06}) and
the class of two components focusing dispersive waves (class of focusing Nonlinear Schr\"odinger equation, \cite{dugrkl09}).
A similar universality property has been also observed in case of the classical dissipative shock (Burgers universality class, \cite{ilin92,duel12}).\\

In the present paper, we consider a fairly general model equation for 1-dimensional scalar unidirectional waves in fluid of the form
\begin{equation}\label{eq:genfluid}
 u_t+a(u)u_x+N[u]=0,
\end{equation}
where $a(u)$ is a non constant function (in most relevant cases $a(u)=u$) and $N$ is a (pseudo)differential operator, generally nonlinear.
Notice that $N$ can be a local operator as well as a nonlocal one, such as the Hilbert transform.
Although our primary interest is the study of dispersive shocks,
this class of equations includes
diffusive and mixed dispersive-diffusive models, which can be selected by different choices of the operator $N$.
More precisely, $N$ models the phenomena into examination by taking into account the relevant physical effects,
like for instance dispersion, dissipation, pressure, or the interfacial interaction between two different fluids.
Notable examples of equations falling into this class include generalized KdV and Burgers equations, the intermediate-long wave and the
Benjamin-Ono equations, and the Benjamin-Bona-Mahony and Camassa-Holm equations. The operators $N$ corresponding to
these equations are listed in Table \ref{tab:exampletable} below. \\
\begin {table}[htpb]
\caption {Notable examples of equations of the form $u_t+a(u)u_x+N[u]=0$.
For every equation in the list we write $a(u)$, $N[u]$, $\bar{N}[u]$, and
the coefficients $\beta$, $\kappa$, $\theta$ of the corresponding universal equation (\ref{eq:linpertcrit}).} \label{tab:exampletable}
\begin{center}
\begin{tabular}{| c | c | c | c | c | c | c |}\hline
\textbf{Equation} & \bf{ $a(u)$} & {\bf $N[u]$} & {\bf $\bar{N}[u]$} & $\beta$ & {\bf$\kappa$} & {\bf $\theta$} \\ \hline
Generalised Burgers & $u^n,\, n>0$ & $-u_{xx}$ & $-u_{xx}$ & $1$ & $1$ & $0$ \\\hline
Generalised KdV & $u^n, \, n>0$ & $u_{xxx}$ & $u_{xxx}$ & $2$ & $0$ & $1$ \\\hline
Benjamin-Ono & $u$ & $-\mathcal{H}[u_{xx}] , \, \Hi[u](x)=\frac{1}{\pi}\dashint \frac{u(y)}{x-y}dy $ & $-\mathcal{H}[u_{xx}]$ & $1$ & $0$ & $1$ \\\hline
Intermediate Long Wave & $u$ & $-\frac{1}{2\delta}\de_x^2\dashint_\mathbb{R}\coth\left(\frac{\pi(x-y)}{2\delta}\right)
u(y)dy +\frac{1}{\delta}u_x $ & $ \frac{\delta}{3}u_{xxx} $ & $2$ & $0$ & $\frac{\delta}{3}$ \\\hline
Camassa-Holm & $u$ & $(1-\de_x^2)^{-1}\left(-uu_x+\frac{2}{3}u_xu_{xx}+\frac{1}{3}uu_{xxx}\right)+uu_x$ &
$-\frac{7}{3}u_x u_{xx}-\frac{2}{3}u u_{xxx}$ & $2$ & $0$ & $-\frac{2}{3} u_c$ \\\hline
Benjamin-Bona-Mahony & $u$ & $(1-\de_x^2)^{-1}\left(-uu_x\right)+uu_x$ & $-3 u_x u_{xx}-u u_{xxx}$ & $2$ & $0$ & $-u_c$ \\\hline
\end{tabular}
\end{center}
\end{table}

The critical behaviour arises
when we consider solutions that at time $t=0$ vary on a large scale (compared to the natural scale of the system), say $1/\e$ with $\e$ small, and
study whether at a later time fluctuations on a smaller scale arise. We assume that the nonlinear operator admits a long-wave expansion,
namely we assume that
there exist a real number $\beta>0$ and an operator $\overline{N}$ such that for any sufficiently smooth function $u$
\begin{equation}\label{eq:longwaveexp}
N S_{\e} [u]=\e^{\beta+1} S_{\e} \overline{N}[u] + o (\e^{\beta+1})\, , \quad \beta >0 \; .
\end{equation}
where $S_\e, \e>0$ is the dilation operator
\begin{equation}\label{dilaton}
 S_{\e}[u](x)=u(\e x) \; .
\end{equation}
In Table \ref{tab:exampletable}, it is shown that assumption (\ref{eq:longwaveexp}) holds for any notable equation listed above.
It follows from (\ref{eq:longwaveexp}) that in the long-wave regime, applying the change of variables 
$(x,t)\to (\tilde{x}\e, \tilde{t} \e)$ and dropping the tildes, the wave satisfies the rescaled equation
$$
u_t+a(u)u_x+\e^{\beta}\overline{N}[u(x)]+o(\e^{\beta})=0 \, , \; \e \to 0 \; ,
$$
with initial data independent on $\e$.
If the latter equation is well-posed, then in this regime (\ref{eq:genfluid}) is a small perturbation of
the scalar conservation law (or Hopf equation)
$u_t+a(u)u_x=0$, at least as long as the wave remains smooth.
However, when the solution of the Hopf equation develops a shock -- a point with vertical derivative --
the term $N[u]$ is not anymore negligible and it makes the wave fluctuate on
a smaller scale, provided the perturbation is not strictly dissipative.\\

In this paper, we analyze -by means of some heuristic techniques reminiscent of \cite{du06}-
the behaviour of waves close to the shock for equations of type
(\ref{eq:genfluid}), and study its universal behaviour.
We show that there is an emerging meso-scale at
which the shock is indeed universal, and described
by a particular solution of the universal PDE
\beq \nonumber
U_T+UU_X+\int_{-\infty}^{+\infty}e^{ipX}\left(\kappa  + i
\theta \sign{p}  \right)|p|^{\beta+1} \hat{U}(p,T)\,dp = 0 \; ,
\eeq
where $ \hat{U}(p,T)$ is the Fourier transform of $U(X,T)$ and
$\kappa,\theta,\beta$ are certain scalar parameters, which can be computed explicitly from the operator $N$. Universality
classes are parametrized by pairs of the form $(\frac{\kappa}{\theta} ,\beta)$; different choices of the parameters give rise to
different universality classes. As a consequence of this classification, we can answer a question posed in \cite{mixu11}:
the Benjamin-Ono equation, although being a Hamiltonian (conservative) PDE,
does not belong to the KdV universality class $(\kappa=0,\beta=2)$ but represents a new universality class $(\kappa=0, \beta=1)$.\\

The second part of the paper is devoted to a characterization of the particular  solution of the universal PDE
that describes the wave at the shock. Our construction is based on the concept of string equation \cite{du06,mara12},
which is a perturbative deformation of the algebraic equation describing the shock of the Hopf equation.
Each universality class is characterized by a particular string equation, and the sought universal solution turns out to
be a specific solution of a boundary value problem for the corresponding string equation. With this approach, we can compute
by perturbative techniques a quantity which is beyond all order in the standard perturbative expansion of solutions of (\ref{eq:genfluid})
in powers of $\e$.
In a few particular cases, the string equation can be computed exactly. For instance,  we recover the string equations obtained for the
KdV and the Burgers classes (an ODE of Painlev\'e type and the Pearcey equation, respectively), and we compute the string equation
for Benjamin-Ono, which was unknown before.
We therefore conjecture that the critical behaviour of solutions to the Benjamin-Ono equation is described
by a particular solution of the singular integro-differential equation
$$X-UT +U^3 -3\,U\Hi[U_X]-3\,\Hi[UU_X]-4\,U_{XX}= 0,$$
where $\mathcal{H}$ is the Hilbert transform on the line. We study the above equation numerically, and we compare our results with
the universal behaviour of KdV and Burgers.  The numerical solution of the Benjamin-Ono string equation requires an effective scheme
for a non-local boundary value problem with an irregular boundary behaviour, and no method was available in the existing  literature.
Our algorithm uses a novel numerical scheme, developed ad-hoc in \cite{olver14},
to effectively evaluate the Hilbert transform for functions slowly-decaying at infinity.\\

Before tackling the precise description of the behaviour of waves at the critical points,  we collect in the next section some known result from
the theory of dispersive shocks. We describe these phenomena, as well as the new results of the present work, using
the language of statistical mechanics. This is done in order to show the deep analogy between critical behaviour of nonlinear waves
and the theory of phase-transitions in statistical mechanics, and with the
purpose to make the dispersive-shock phenomenon more easily achievable by any scientist familiar with the latter theory. In doing so
we will unfold a dictionary between the critical behaviour of nonlinear waves and the phase transitions of statistical mechanics,
summarised in Table \ref{tab:onlytable} \footnote{It is interesting to compare our Table with Table 1 of
\cite{mor13} concerning the dissipative shock.}.
\begin {table}[htpb]
\caption {Dictionary: Nonlinear Waves and Statistical Mechanics} \label{tab:onlytable} 
\begin{center}
\begin{tabular}{|l|l|}\hline
Wave equation in 1+1 dimensions & Statistical Model\\\hline
Long-wave limit & Thermodynamic Limit\\\hline
Critical Point of Gradient Catastrophe & Tricritical Point\\\hline
Wave Amplitude in the Whitham Zone & Order Parameter\\\hline
Unfolding of Cubic Singularity & Mean Field of $\varphi^6$ model\\\hline
Meso-scale at the critical point & Renormalization group flow\\\hline
Scaling Linear perturbations of Hopf & Fixed point of renormalization group\\\hline
\end{tabular}
\end{center}
\end{table}

\section{Dispersive shock as a tricritical phase transition}
\begin{figure}[htpb]
  \centering
  \subfigure[Profile of the typical solution of KdV in the semiclassical regime (here $\e=10^{-2}$) after the critical time.
  Figure from the arXiv version of \cite{grkl07}.
  \label{fig:smalle2}]{\includegraphics[width=5cm,height=3.5cm]{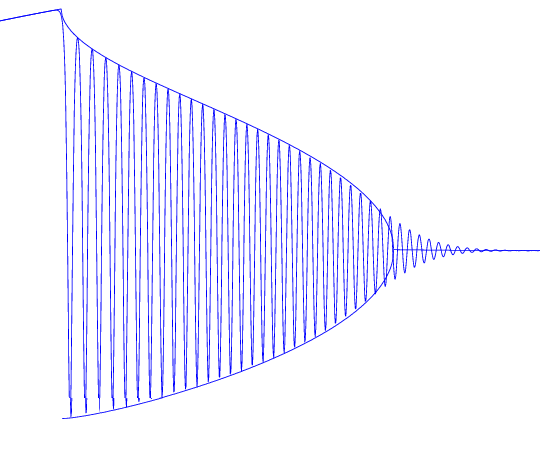}} \hfill
  \subfigure[Profile of the amplitude of oscillations $W$ for a typical solution of KdV  and its mean-field approximation $W_{MF}$.
   $W$ is discontinuous at the left boundary of the shock region and continuous but not differentiable
  at the right boundary.\label{fig:smalle4}]{\includegraphics[width=5cm,height=3.5cm]{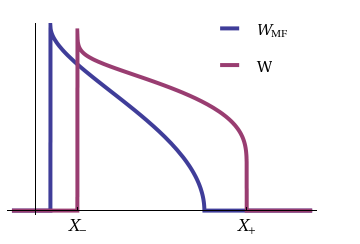}} \hfill
  \subfigure[Phase portrait of the typical solution in the $(x,t)$. Straight lines are the characteristics lines. The thick line
  is the boundary of the shock region $W>0$ (or Whitham zone). In general it does not
  coincide exactly with region where characteristics line interesect.
  \label{fig:smalle1}]{\includegraphics[width=5.5cm,height=3.5cm]{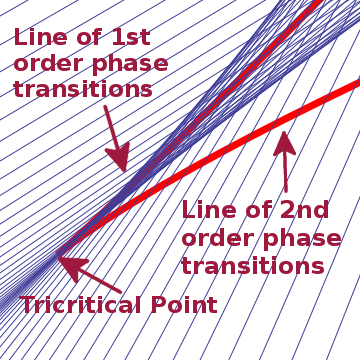}}
\end{figure}

To understand the dispersive-shock phase-transition, we look at the well understood case of the dispersionless (or semiclassical)
limit of Korteweg-de Vries (KdV) \cite{gupi73,lale83, deift98}
\begin{equation*}
 u_t+uu_x-\e^2 u_{xxx}=0 \ , \e \to 0 \, , \; u(x,t=0,\e)=\varphi(x).
\end{equation*}
We assume the initial data to be smooth, positive, fast decaying and with a single hump. The formal $\e=0$ limit,
known as Hopf equation, describes a wave where every particle on the profile
travels with constant velocity $u$, i.e. the solution is constant along the characteristic lines
\begin{equation}\label{eq:charlines}
 x(t;x_0)=x_0+\varphi(x_0) t  \; , \qquad u(x(t;x_0),t)=\varphi(x_0).
\end{equation}
Up to the critical time $t_c$ all lines are distinct and the solution is uniquely determined, while after it the lines
start to intersect and the wave develops a shock, a point with vertical derivative, where the contribution $\e^2 u_{xxx}$ is
not negligible, no matter how small $\e$ is. Lax and Levermore \cite{lale83} showed that in the semiclassical regime the $(x,t)-$plane
is divided in two zones (see Figure \ref{fig:smalle1}). In the first, known as the \emph{semiclassical zone} and containing the strip
$\mathbb{R}\times [0,t_{c})$, the limit $\lim_{\e \to 0}u(x,t,\e)=u(x,t)$ exists and corresponds to the solution of the Hopf equation,
or one of its branches if it is multi-valued.
In the second, known as \emph{Whitham zone},
$u(x,t,\e)$ develops oscillations of vanishing wavelength $O(\e)$ (see Figure \ref{fig:smalle2}) and
the limit exists only in a weak sense:
there exists a function $\overline{u}(x,t)$
which averages the oscillations, uniquely defined by the weak limit
$$\lim_{\e \to 0}\int \psi(x)u(x,t,\e) dx = \int \psi(x) \bar{u}(x,t) dx,$$
for any test function $\psi(x)$ \footnote{For times sufficiently big after the time of the shock $t_c$,
the solution inside the Whitham zone may undergo further shocks and the Whitham
zone is thus divided in subregions of different 'genera' \cite{lale83}.
These additional shocks are outside the scope of the present paper, which is concerned with the behaviour of solutions for $t \sim t_c$.}.
The boundary of the Whitham zone depends on the initial data only;
an approximate expression of the boundary, up to a certain time beyond the critical one and for certain classes of initial data,
has been described by Grava and Klein \cite{grkl07}. To better understand the transition of the solution from the regular to the
oscillatory behaviour, we introduce an \textbf{order parameter} $W(x,t)$, which measures the amplitude of the oscillations in the Whitham zone:
\begin{equation}\label{eq:orderpar}
 W(x,t)=\lim_{\e \to 0} \sup |u(x,t,\e)-\overline{u}(x,t)|  \; .
\end{equation}
In the KdV case the function $W$, shown in Figure \ref{fig:smalle4}, can be computed exactly from the formulas for $u$ and $\bar{u}$ obtained
in \cite{grkl07}.
Let us fix $t$ at a value bigger than $t_c$. The Whitham zone is then an interval $(x_-(t),x_+(t))$ of the real line.
The order parameter $W(x,t)$ is zero outside that interval, it behaves like $W(x,t)
\sim 1/\log(x_+(t)-x)$ close to the right boundary,
while it is discontinuous at the left boundary: $\lim_{x \downarrow x_-}W(x,t)>0$.
Therefore, the solution undergoes a second order phase transition at $x=x_+$ -- the order parameter is continuous but not differentiable --
and a first order phase transition at $x=x_-$ -- the order parameter is discontinuous. The boundary of the Whitham zone
is made of a curve of second order phase transitions and a curve of first order phase transitions, which meet at a point $(x_c,t_c)$,
which is therefore a \textbf{tricritical point}, according to the standard theory of phase transitions in statistical mechanics \cite{landau58}
(see Figure \ref{fig:smalle1}) \footnote{ In the case of Benjamin-Ono the Whitham zone coincides with the region
of the $(x,t)$ plane where the solution of Hopf is multi-valued
\cite{mixu11}. In the
general case, as for instance for KdV, this is not true anymore.}.\\

In what follows we investigate the local behaviour of solutions close to the tricritical point for a general PDE \eqref{eq:genfluid},
we argue that it is universal and we characterize the universality classes. To avoid a cumbersome notation and since the final result
is independent of $a$, we stick with $a(u)=u$ \footnote{The few modifications needed in formulas (\ref{eq:Xchange},\ref{expucrit}) can
be found in \cite{du06}}. For the Hopf equation, the critical point $x_c,t_c$ is the point where the wave breaks and the solution becomes
multivalued, a singular behavior known as
\textbf{gradient catastrophe}.
It is well known that the generic singularity is a cubic one. Indeed, it follows from (\ref{eq:charlines}) that solutions can locally be
expressed by the implicit formula
\begin{equation*}
u(x-\varphi(x)t)=\varphi(x) \quad\mbox{or}\quad x-u\,t=f(u),\; f=\varphi^{-1};
\end{equation*}
if we let $u_c=u(x_c,t_c)$ and suppose
$f'''(u_c) \neq 0 $, then we can introduce the scale variables
\begin{align}\nonumber
&X=\frac{x-x_{c}-u_{c}(t-t_{c})}{\lambda},\quad  U=\left(\frac{\gamma}{6}\right)^{1/3}\frac{u-u_c}{\lambda^{\frac{1}{3}}},\\
&T=\left(\frac{6}{\gamma}\right)^{1/3}\frac{(t-t_{c})}{\lambda^{2/3}}, \label{eq:Xchange}
\end{align}
with $ \gamma=- f'''(u_c)> 0$ and $\lambda$ a small parameter, to get for small $\lambda \to 0^+$
\begin{equation}\label{eq:cubic}
X-U\,T+U^{3}=0 \; ,
\end{equation}
which is the miniversal unfolding of the cubic singularity. A good picture of the transition can be derived from (\ref{eq:cubic}).
If $\beta_1\geq\beta_2\geq\beta_3$ are the three roots of $U$ for $T>0$, then $W_{MF}(x,t)=2(\beta_1(x,t)-\beta_2(x,t))$ measures
the envelope of the solution.
This has the same phase diagram and qualitative behaviour of the exact order parameter KdV, but a different exponent
$W_{MF} \sim (\tilde{x}_{+}(t)-x)^{1/2}$. This \textit{Mean Field} description of the phase transition
gives the correct behavior but discards the true nature of the oscillations.
It shows the same structure as the $\varphi^6$ mean field theory \cite{landau58}.

\section{The universal model PDE at the tricritical point}

Generalizing a procedure described in \cite{du06},
we are able to give for quite a general perturbation $N$ a much finer description of the tri-critical phase transition,
which takes into account the precise
nature of the oscillations and which is remarkably \textbf{universal}. In fact, we argue that the local behaviour of $u$ around
the tricritical point is uniquely characterized by the linearization of $\bar{N}$ at the constant
function $u \equiv u_c$:
$$
\overline{N}[u_c+\delta u]= \overline{L}_{u_c}[\delta u] + O((\delta u)^2) \; ,
$$
where $\overline{N}[u_c]=0$ because of (\ref{eq:longwaveexp}).
More precisely, the idea is that -- at the appropriate scale -- the wave $u$ satisfies a distinguished universal solution
to a model PDE which is uniquely determined by
$\overline{L}$.
% Note that the scaling assumption (\ref{eq:longwaveexp}) on $N$ implies that
% \begin{equation}\label{eq:nbarscales}
% \bar{N} S_{\e}[u]= \e^{\beta+1}S_{\e} \bar{N}[u],
% \end{equation}
% for any sufficiently smooth function $u$; here $S_{\e}$ is the dilation operator defined in (\ref{dilaton}). The identity (\ref{eq:nbarscales})
% is proved in the Appendix.
% Before we proceed with the discussion of the universal PDE we make the assumption that
% for sake of definiteness, we restrict ourselves to those operators $N$ such that $\overline{L}_{u_c} $ is a pseudo-differential operator
% for any constant function $u \equiv u_c$.
% We remind that the class of pseudo-differential operators encompasses differential operators,
% classical integral operators and any combination of them \cite{stein93}.
% that $\bar{N}$ (and therefore $\bar{L}_c$) is translationally invariant \footnote{The case
% of a non-translationally invariant $\bar{N}$ can be theoretically dealt with our methods but it seems to be less relevant for studying the long-wave limit
% of unidirectional waves.}.
%
The aim of this section is to derive and classify the universal PDEs;  in Section \ref{sec:string} we will consider
the characterization of the particular solution. To set the appropriate scale around the tri-critical point,
let us change variables as in (\ref{eq:Xchange}) but with the scaling parameter $\lambda=\e^{\frac{1}{\alpha}}$ depending on $\e$.
A simple computation shows that the wave equation (\ref{eq:genfluid}) reduces to
\begin{equation*}
 U_T+UU_X+\e^{\frac{\beta(\alpha-1)-1/3}{\alpha}}\,\bar{L}_{u_c}[U]+ \mbox{ higher order terms} =0.
\end{equation*}
The balance of the intrinsic and extrinsic scales $\e,\lambda$ is achieved when $\alpha=1+\frac{1}{3 \beta}$,
or equivalently if $u$ admits the expansion
\begin{align}
&u(x,t,\e)\simeq u_{c}
+\e^{\frac{ \beta }{3 \beta + 1}}\, U\left(\frac{x-x_{c}-u_{c}(t-t_{c})}{\e^{\frac{3 \beta }{3 \beta +1}}},
\frac{t-t_{c}}{\e^{\frac{2 \beta }{3 \beta + 1}}}\right)\notag\\ 
&+O(\e^{\frac{2 \beta}{3\beta+1}}) , \label{expucrit}
\end{align}
and the leading term $U$  of \eqref{expucrit} is a solution of the linearized perturbation
\begin{equation}\label{eq:critlinear}
 U_T+UU_X+\bar{L}_{u_c}[U]=0.
\end{equation}
The universal behavior close to the tricritical point emerges thus on a \textbf{meso-scale $\e^{1/\alpha}$}, lying between the microscopic $O(\e)$ scale
and the macroscopic one $O(\e^0)$.
The reader should compare this situation with the case of renormalization group in statistical mechanics,
where universality arises by \textit{magnifying} the theory at the meso-scale where block spin or phase-space renormalization is performed \cite{wilson71}.\\

Before describing the \textbf{distinguished solution} $U(X,T)$ of (\ref{eq:critlinear}) which gives the universal correction at the tricritical point,
we consider the classification of Universality Classes. Since any positive constant in
front of $\bar{L}_{u_c}$ can be factored out trivially, we say that \emph{two nonlinear PDEs $N, N'$
belong to the same universality class if $\bar{L}_{u_c}=\bar{L'}_{u_c}$ up to a (positive) scalar multiple}.
The classification of \textbf{universality classes} coincides with the classification of linear operators $\bar{L}_c[U]$ that are the linearization of an operator $N$
admitting the long-wave expansion (\ref{eq:longwaveexp}). By assumption on $N$,
$\bar{L}_c$ is a linear pseudo-differential operator. If we further assume that 
$L_c$ is traslationally invariant \footnote{The case
of a non-translationally invariant $\bar{N}$ can be theoretically dealt with our methods but it seems to be less relevant for studying the long-wave limit
of unidirectional waves.}, then it admits the representation
$L[U](x):=\int_{-\infty}^{+\infty}e^{ipx}m(p)\hat{U}(p)\,dp$ for some regular enough
Fourier multiplier $m(p)$ \cite{stein93}. Here $ \hat{U}(p)$ stands for the Fourier transform of $U$. In order for the operator to define a meaningful evolution,
it must map real functions into real functions and it must be either conservative or dissipative, i.e. $\int U(X)L[U(X)]dX \geq 0$; the two conditions read $m(-p)=m^*(p)$, $Re(m(p))\geq 0.$ A complete characterization of the admissible operators $\bar{L}$  is achieved by means of the following fact which is proved in the Appendix: the scaling assumption (\ref{eq:longwaveexp}) on $N$ implies that
\begin{equation}\label{eq:nbarscales}
\bar{N} S_{\e}[u]= \e^{\beta+1}S_{\e} \bar{N}[u],
\end{equation}
for any sufficiently smooth function $u$, where $S_{\e}$ is the dilation operator defined in (\ref{dilaton}). Due to (\ref{eq:nbarscales}), then $\bar{L}$ satisfies the  same scaling law
$$\bar{L}_{u_c} \circ S_{\e}=\e^{\beta+1} S_{\e} \circ \bar{L}_{u_c} ,$$
and this  further constrains the Fourier multiplier to be of the form
$m(p)=\kappa |p|^{\beta+1}+i \theta\, p\, |p|^{\beta} \; ,$
for some $(\kappa,\theta)\in \bb{R}^2\setminus\{0\}$, $\kappa\geq0$, and $\beta>0$.  Explicitly, we have:
\beq\label{eq:linpertcrit}
U_T+UU_X+\int_{-\infty}^{+\infty}e^{ipX}\left(\kappa  + i
\theta \sign{p}  \right)|p|^{\beta+1} \hat{U}(p)\,dp = 0 \; .
\eeq
Critical universality classes are thus characterized by a pair of parameters $(\frac{\kappa}{\theta}, \beta)$.
Since the transformation $U(X,T)\to -U(-X,T)$
sends $\theta$ to $-\theta$, we can assume $\theta \geq 0$.
Notice that if $\theta=0$ then the perturbation is purely dissipative while if $\kappa=0$ it is dispersive and possesses the Hamiltonian
$H[U]=\int_{-\infty}^{+\infty} U^3-\theta \,U\,K[U]dX,$
where $K[U]=\int e^{ipX}|p|^{\beta} \hat{U}(p)\,dp.$

\begin{exa} Conservation laws $u_t+\partial_x f(u,u_x,\dots)=0$, with $f$ some smooth function,  admit a long-wave regime with
$\bar{N}[u(x)]=\partial_x(n(u)u_x)$, where $n(u)=\frac{\partial f}{\partial u_x(x)}_{|u_x=u_{xx}=\dots =0}$. Provided $n(u_c) \neq 0$, then $\beta=1$ and
$\bar{L}_{u_{c}}=n(u_{c})\,u_{xx}$. The universal model for these equations is thus the Burgers equation:
$$U_{t}+U\,U_{x}+n(u_{c})\,U_{xx}=0.$$
The critical behaviour for these conservation laws is typical of dissipative shocks and it has been considered in \cite{duel12,arlomo13}.
\end{exa}
\begin{exa} Local Hamiltonian PDEs are equations in the form
$u_t=\du{} \int h(u,u_x,\dots) dx$, for a smooth function $h$ s.t. $h(0,0,\dots)=0$. They admit a long-wave expansions
with $\bar{N}[u]=\de_{x}(b'(u)\,u_{x}^{2}+2b(u)\,u_{xx})$, $b(u)=\frac{\partial h}{\partial u_x(x)}_{|u_x=u_{xx}=\dots =0}$.
Provided $b(u_c) \neq 0$, then $\beta=2$ and  $\bar{L}_{u_{c}}=b(u_{c})\,u_{xxx}$. Therefore the universal model for these equations is KdV:
$$U_{t}+U\,U_{x}+b(u_{c})\,U_{xxx}=0.$$
The critical behaviour for this class has been considered in \cite{du06}.
\end{exa}
\begin{exa} The Benjamin-Ono (B-O) equation \cite{ben67,ono75} 
$$u_{t}+u\,u_{x}- \mathcal{H}[u_{xx}]=0,$$
where  $\Hi$ is the Hilbert transform: $\Hi[u](x)=\frac{1}{\pi}\dashint \frac{u(y)}{x-y}dy$ is an integrable Hamiltonian equation as KdV,
but it is non-local. The operator $\mathcal{H}[u_{xx}]$ is a translationally-invariant
pseudo-differential operator with Fourier multiplier $m(p)=i \sign{p} p^2$ \cite{stein93}. Thus, in this case $N=\bar{L}=\Hi[u_{xx}]$. Therefore
the Benjamin Ono equation is already in the long-wave form \eqref{eq:linpertcrit}, with $\beta=1$, $\kappa=0$ and $\theta=1$. 
It has the same exponent as Burgers
but being Hamiltonian like KdV its solutions undergo a dispersive shock \cite{mixu11}. It
corresponds therefore to a novel universality class, which we name \emph{Benjamin-Ono universality class}.
All equations of B-O hierarchy (see \cite{mixu12} for the definition) belong to the B-O universality class.
\end{exa}
\begin{exa} The Intermediate Long Wave (ILW) equation \cite{jos77}
$$u_{t}+u\,u_{x}+\frac{1}{\delta}\,u_x+\mathcal{T}_\delta[u_{xx}]=0,$$
where
$$\mathcal{T}_\delta[u(x)]=-\frac{1}{2\delta}\dashint_\mathbb{R}\coth\left(\frac{\pi(x-\xi)}{2\delta}\right) u(\xi)d\xi,$$
and $\delta\in\mathbb{R}$, is an integrable equation which models nonlinear waves in a fluid of finite depth. Moreover, in the limit $\delta\to 0$ one formally gets the KdV equation, while the limit $\delta \to \infty$ gives the Benjamin-Ono equation. It is therefore interesting to check whether this equation belongs to one of the above mentioned universality classes, or possibly to a new one. From the representation $\mathcal{T}_\delta[u(x)]=i \int_{-\infty}^{+\infty}e^{ipx}\coth(\delta p)\hat{u}(p)\,dp$ (see \cite{hono04}), together with the expansion $\coth(\delta p)=\frac{1}{\delta p}+\frac{\delta}{3}p+O(p^3)$,  it follows that -- as long as $\delta$ remains finite -- the ILW equation belongs to the KdV universality class.
\end{exa}
\begin{exa} The Camassa-Holm equation \cite{caho93}:
$$u_{t}-u_{xxt}+uu_{x}=\frac{2}{3}u_{x}u_{xx}+\frac{1}{3}uu_{xxx},$$
and
the Benjamin-Bona-Mahony equation \cite{beboma72}
$$u_{t}-u_{xxt}+uu_{x}=0$$
can be written in the standard form (\ref{eq:genfluid}) by inverting $1-\partial_x^2$.
Provided $u_c \neq 0$, they also belong to the KdV universality class \cite{du06}.
\end{exa}
We note that the universal PDE \eqref{eq:linpertcrit} is again of the form of equation \eqref{eq:genfluid}.
In many important special cases, such as KdV, Burgers and Benjamin-Ono, the procedure of rescaling at the tricritical point
simply reproduces the equation that one started with. In other words, the special case $a(u)=u, N = \bar{L}_{u_c}$ is \textbf{invariant} under rescaling.
This explains the naming convention of the universality classes appearing in the examples above. A comparison between the different
universality classes for the examples considered above can be found in Table \ref{tab:exampletable}

\section{The Universal Correction as solution of the String Equation}\label{sec:string}
In this section, we show how to compute the universal correction
$U(X,T)$, defined by
the multiscale expansion (\ref{expucrit}) of $u$ at the tricritical point,
as a particular solution of the universal model equation
(\ref{eq:critlinear}).
We argue that $U$ is the solution of a (possibly infinite) deformation
of the cubic equation (\ref{eq:cubic}), known as \textbf{string equation}.
Our approach generalizes and simplifies the one originally proposed in
\cite{du06} for the Hamiltonian case, and developed
mathematically by the authors \cite{mara11,mara12} (see also
\cite{arlomo13} for the Burgers universality class). We derive the string equation using a simple principle,
rigorously proved in \cite{mara12} in some generality: \textit{in the long-wave regime, any solution of equation
\eqref{eq:genfluid} - and in particular of equation \eqref{eq:critlinear} - can be uniquely characterized as the fixed
point of a symmetry. This symmetry arises as deformation of a symmetry of the Hopf equation.} This principle can be applied
to characterize the universal correction $U(X,T)$.
First, we note that the rescaled function
$U^{\mu}(X,T)=\mu^{3\beta+1}U(X/\mu^{9 \beta+3},T/\mu^{\frac{6
\beta+2}{3}})$ satisfies the
long-wave limit of (\ref{eq:critlinear}), namely
\beq\label{eq:longlimitcrit}
U^{\mu}_T=U^{\mu}U^{\mu}_X+\mu\, \bar{L}_{u_c}[U^{\mu}].
\eeq
The limit $\mu \to 0$ is well-defined, for $U^{0}(X,T)$ coincides with
the solution of the cubic equation (\ref{eq:cubic}). In addition,  the latter is a 
solution of the Hopf equation that can be characterized as the unique stationary solution (vanishing at $X=T=0$) of the flow
\begin{equation}\label{eq:symHopf}
 U^{0}_S=\partial_X(X-U^{0}\,T+\,(U^{0})^{3}),
\end{equation}
which commutes with Hopf. In other words, $U^0$ is the
fixed point of the symmetry generated by the flow \eqref{eq:symHopf}. We now follow the general principle stated
above to characterize $U^{\mu}(X,T)$ as the fixed point of the flow
\begin{equation}\label{eq:deformedsymmetry}
U^{\mu}_{S}= \partial_X(X-U^{\mu}\,T+(U^{\mu})^3 +\mu\, \alpha_1[U^{\mu}]+\mu^2 \alpha_2[U^{\mu}]+ \dots),
\end{equation}
obtained as the (unique) power series in $\mu$ commuting order by order with \eqref{eq:longlimitcrit} \footnote{The
existence of such a deformation can be established in case $L$ is a
differential operator
by the method of \cite{lizh06}.}.
By definition, the \textbf{string equation} is the equation for the
vanishing of the right hand-side
of (\ref{eq:deformedsymmetry}).
In general, we expect the symmetry to be an infinite (possibly not
converging) power series in $\mu$; in that case the string equation
will be valid
only asymptotically for $\mu$ small, or equivalently for $X \gg 0$. However, if the string equation truncates, we can safely put
$\mu=1$ to get the exact form of $U \equiv U^{1}$. 
We conjecture that the function $U(X,T)$ is then uniquely
characterized as the solution of the string equation satisfying the boundary
behaviour
\begin{equation}\label{eq:UXgrande}
 U(X,T) \sim - \sign{X}|X|^{\frac{1}{3}}\; \text{ as }\; |X|
\to \infty,\; \forall \,T,
\end{equation}
which assures the correct long-wave ($\mu \to 0$) limit.
The string equation is finite for at least three universality classes: Burgers, Benjamin-Ono and KdV. We point out that the method of the string equation
is a valid alternative to the classical approach to shock by means of
step-function initial data \cite{gupi73}, for it
contains all universal information.

\begin{figure}[htpb]
  \centering
  \subfigure[Solution of the cubic equation, representing solutions of the Hopf equation close to the shock. After the shock $T>0$
  the solution is multi-valued as the Hopf equation is not anymore a good model\label{fig:hopf}]
  {\includegraphics[width=8cm,height=5.5cm]{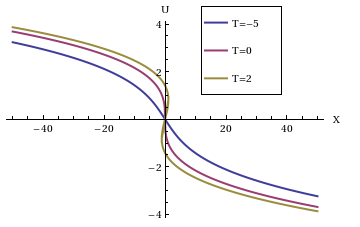}}\hfill
\subfigure[Solution of eq. \eqref{eq:buP12}, representing the universal transition from a regular wave ($T\!<\!0$) to
a classical shock wave ($T\!>\!0$). As expected, the wave steepens but no oscillations arise.
 \label{fig:burgers}]{\includegraphics[width=8cm,height=5.5cm]{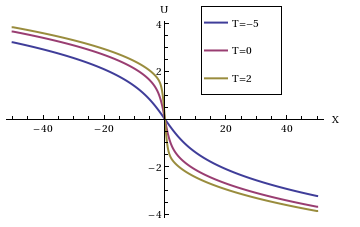}}\hfill
\subfigure[Solution of KdV string equation, representing solutions of equations in KdV class close to the shock. For $T>0$
the wave oscillates showing the typical pattern of a dispersive-shock.  \label{fig:kdv}]
{\includegraphics[width=8cm,height=5.5cm]{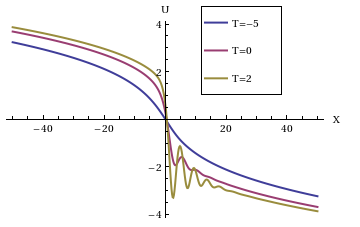}} \hfill
 \subfigure[Solution of B-O string equation. The shock is again dispersive but the oscillations appear to have shorter wave-length and bigger amplitude
  with respect to the KdV case. \label{fig:bono}]
  {\includegraphics[width=8cm,height=5.5cm]{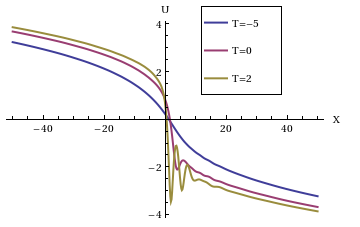}}
\end{figure}

\begin{exa}
Burgers universality class. The equation
$$U_S=\partial_X(X-UT+U^{3}-6UU_{X}+4\,U_{XX}),$$ 
is a symmetry of Burgers (see \cite{arlomo13}) and the string equation is therefore
\begin{equation}\label{eq:buP12}
X-UT+U^{3}-6UU_{X}+4\,U_{XX}=0.
\end{equation}
The unique solution satisfying \eqref{eq:UXgrande} can be explicitly written in terms of the Pearcey integral \cite{duel12,arlomo13},
and it is plotted in Fig \ref{fig:burgers}. It is proved \cite{ilin92} that sufficiently regular solutions of the
Burgers equation admits an expansion (\ref{expucrit}) where $U(X,T)$ is exactly the solution of (\ref{eq:buP12}).
\end{exa}
\begin{exa}
The KdV universality class. The equation 
$$U_S=\partial_X(X-U\,T+U^{3}-3\,U_{X}^{2}-6\,U\,U_{XX}+\frac{18}{5}\,U_{XXXX})$$ 
is a symmetry of  the KdV  equation \cite{du06}, and the string equation satisfied by $U(X,T)$ is
\begin{equation}\label{eq:P12}
X-U\,T+U^{3}-3\,U_{X}^{2}-6\,U\,U_{XX}+\frac{18}{5}\,U_{XXXX}= 0.
\end{equation}
The unique solution satisfying the boundary condition (\ref{eq:UXgrande}) \cite{clva07} is plotted in Fig \ref{fig:kdv}.
This ODE is known in literature as the second equation of the Painlev\'e I hierarchy and appeared in the context of random matrix theory
\cite{douglas90, brezin90}. It is known that Painlev\'e equations can be linearized by means of an isomonodromy system \cite{fokas06}.
It was proved in \cite{claeys2009,clgr11} that sufficiently regular solutions of any equation of the KdV hierarchy admits an expansion (\ref{expucrit})
where $U(X,T)$ is exactly the solution of
(\ref{eq:P12}). The extension of this result to other local Hamiltonian PDEs
-- yet to be proved -- goes under the name of Dubrovin's universality conjecture \cite{du06}.
\end{exa}
\begin{exa}
The Benjamin-Ono String equation for $U$ is finite and given by the formula 
\begin{equation}\label{eq:boP12}
X-UT +U^3 -3\,U\Hi[U_X]-3\,\Hi[UU_X]-4\,U_{XX}= 0,
\end{equation}
as both $\partial_X(X-UT)$ and $\partial_X(U^3 -3\,U\Hi[U_X]-3\,\Hi[UU_X]-4\,U_{XX})$  are symmetries of B-O \cite{mixu12}.
\end{exa}

\paragraph*{A new Painlev\'e equation?}
Equation \eqref{eq:boP12} is particularly important, as it is the first time that a nonlocal ODE resembling a
Painlev\'e equation appears in the literature. We investigated equation \eqref{eq:boP12} numerically using a spectral method where
the Hilbert transform is computed following \cite{olver14}.
According to our numerical results - see Fig \ref{fig:bono} - for any real $T$ equation \eqref{eq:boP12} admits a
unique solution satisfying (\ref{eq:UXgrande}) and such $U$ solves Benjamin-Ono. We point out that equation (\ref{eq:boP12})
is a candidate for a new class of Painlev\'e equations. In fact, both
\eqref{eq:buP12} and \eqref{eq:P12} can be linearized and satisfy the Painlev\'e property \cite{fokas06},
for any solution extends to a meromorphic function in the complex plane.
Here various important questions arise:
does the unique solution of (\ref{eq:boP12})  satisfying (\ref{eq:UXgrande}) expand to a meromorphic function?
Does (\ref{eq:boP12}) admit a linearization by means of
an isomonodromic system? We remark that in many cases the isomondromic system for a string equation arising from an integrable hierarchy
can be constructed from the zero-curvature representation of the latter \cite{flas80,mo90}. However, to the best of our
knowledge, no zero-curvature representation of Benjamin-Ono has been so far discovered.  \\

We conclude this letter summarizing our results. By introducing an
order parameter -- the wave amplitude in the Whitham zone -- we
analyzed the dispersive-shock-transition by means of statistical
physics and we showed that it corresponds to a tri-critical point.
Generalizing an argument by Dubrovin \cite{du06},
we then refined the coarse description
of the transition using the amplitude of oscillations in the Whitham
zone and we unveiled
the precise local behaviour of the wave close the shock by means of
the string equation, which encodes all universal
features of the transition. In particular, we obtained the explicit
description of the critical behaviour of solutions
to the Benjamin-Ono equation, which we had showed to be the model
equation for a new universality class of
dispersive shock waves.
Albeit our interest is primarily the dispersive-shock, our
classification of universality classes
does comprehend also dissipative and dispersive-dissipative equations
alike, modeling media where dispersion
and diffusion are balanced.
Interestingly, their universal classes are
PDEs with non-local interactions, similar to
those used in experimental realization of dispersive shocks \cite{cfprt09}.
We are currently building the necessary mathematical technology
to explicitly compute the string equation for these more general cases.
It would be very interesting to rigorously prove the results claimed in the present paper,
particularly concerning the critical behaviour of the Benjamin-Ono equation.\\

\noindent
\emph{Acknowledgments} We thank Peter Miller, Nelson Bernardino and Antonio Moro for useful comments,
and Sheehan Olver for his suggestions about the numerical approach to the Hilbert transform.
D. M. is supported by the FCT Post Doc Fellowship number SFRH/BPD/75908/2011.
P. A. is supported by the FCT Investigator Grant IF/00177/2013/CP1159/CT0009.
D.M and A. R. are partially supported by GNFM through the program Progetto Giovani.

\section{Appendix}
\noindent
We prove formula (\ref{eq:nbarscales}). We suppose that $N$ is a continuos operator on the space of Schwartz functions, which can possibly
be extended to a larger space by continuity. We claim that if there exist a $\beta>0$ and a second operator $\bar{N}$ such that $N$ satisfies
\begin{equation*}\label{eq:Ndilated}
     S_{\e^{-1}}NS_{\e}=\e^{\beta+1} \bar{N}+ o(\e^{\beta}),
    \end{equation*}
for any  Schwartz function $u$, then
\begin{equation*}\label{eq:Nbardilated}
     S_{\e^{-1}}\bar{N}S_{\e}=\e^{\beta+1} \bar{N} ,
\end{equation*}
which is precisely (\ref{eq:nbarscales}). The dilation operator $S_{\e}$ appearing above is defined as in \eqref{dilaton}.
To prove our claim, we notice that the dilation operator is a continuous operator on the space of Schwartz functions. Therefore it acts continuously, by conjugation,
on the space of continuous operators on that space (embedded with the weak topology). Therefore, the thesis follows from the following:
\begin{lem}\label{lem:semigroup}
 Let $G(\e)$, $\e >0$, be a group of continuous operators on a vectors space $V$. If $G(\e)v=\e^{\alpha}\bar{v}+ o(\e^{\alpha})$ then
 $G(\e)\bar{v}=\e^{\alpha}\bar{v}$.
 \end{lem}
\emph{Proof.}  $\bar{v}=\lim_{\e \to 0} \frac{G(\e)v}{\e^{\alpha}}$. Since $G$ acts continuously then for every $\delta\in \bb{R}^+$,
 $G(\delta)\bar{v}=\lim_{\e \to 0} \frac{G(\delta \e)v}{\e^{\alpha}}=\delta^\alpha \bar{v}$. Q.E.D.

\def\cprime{$'$} \def\cprime{$'$} \def\cprime{$'$} \def\cprime{$'$}
  \def\cprime{$'$} \def\cprime{$'$} \def\cprime{$'$} \def\cprime{$'$}
  \def\cprime{$'$} \def\cprime{$'$} \def\cydot{\leavevmode\raise.4ex\hbox{.}}
  \def\cprime{$'$} \def\cprime{$'$} \def\cprime{$'$}

%\bibliographystyle{plain}
%\bibliographystyle{alpha}
%\bibliographystyle{unsrt}
%\bibliography{biblio}

\end{document}